\documentclass[prd,floatfix,preprintnumbers, superscriptaddress]{revtex4}
\usepackage{amsmath}
\usepackage{natbib}
\usepackage[final]{graphicx,epsfig}
\setcitestyle{numbers,square,sort&compress}
\usepackage{url}
\usepackage{subfig}

\newcommand{\bfl}{\begin{flushleft}}
\newcommand{\efl}{\end{flushleft}}
\newcommand{\bea}{\begin{eqnarray}}
\newcommand{\eea}{\end{eqnarray}}
\newcommand{\be}{\begin{equation}}
\newcommand{\ee}{\end{equation}}
\newcommand{\bi}{\begin{itemize}}
\newcommand{\ei}{\end{itemize}}

\def\bec{\begin{center}}
\def\eec{\end{center}}
\def\beq{\begin{equation}}
\def\eeq{\end{equation}}

\begin{document}

\title{Cosmology with Independently Varying Neutrino Temperature and Number}
\author{Richard Galvez}
\affiliation{Department of Physics and Astronomy, Vanderbilt University, Nashville, TN 37212, USA}
\affiliation{Department of Physics and Astronomy, Fisk University, Nashville, TN 37208, USA}
\author{Robert J. Scherrer}
\affiliation{Department of Physics and Astronomy, Vanderbilt University, Nashville, TN 37212, USA}
\email{Richard.A.Galvez@Vanderbilt.edu} 
\email{Robert.Scherrer@Vanderbilt.edu}

\begin{abstract}
We consider Big Bang nucleosynthesis and the cosmic microwave background
when both the
neutrino temperature and neutrino number are allowed to vary from their standard values.
The neutrino temperature is assumed to differ from its standard model value by a fixed factor
from Big Bang nucleosynthesis up to the present.
In this scenario, the effective number of relativistic
degrees of freedom, $N_{\rm eff}^{\rm CMB}$, derived from observations of
the cosmic microwave background is not equal to
the true number of neutrinos, $N_\nu$.
We determine the element abundances
predicted by Big Bang nucleosynthesis as a function of the neutrino
number and temperature, converting the latter to the equivalent value
of $N_{\rm eff}^{\rm CMB}$.  We find that a value of $N_{\rm eff}^{\rm CMB} \approx 3$ can
be made consistent with $N_\nu = 4$ with a decrease in the neutrino temperature of $\sim 5\%$,
while $N_\nu = 5$ is excluded for any value of $N_{\rm eff}^{\rm CMB}$.  No observationally-allowed
values for $N_{\rm eff}^{\rm CMB}$ and $N_\nu$ can solve the lithium problem.

\end{abstract}

\maketitle

\section{Introduction}

Cosmological observations currently provide some of
the most informative probes of physics beyond the Standard Model of particle physics.
An analysis of the temperature anisotropies from the Cosmic Microwave Background (CMB) and the predictions made by Big Bang Nucleosynthesis (BBN) are two of the most robust methods available to gain an understanding of the physics governing the early Universe.

One of the first such cosmological constraints drawn from such an analysis was the derivation of an upper limit on the number of neutrinos inferred from the abundance of $^4$He using BBN predictions to be $N_\nu \leq 5$ \cite{Steigman:1977kc}. More recently, precise measurements of the CMB by the WMAP \cite{Hinshaw:2012aka} and PLANCK \cite{Ade:2015xua} collaborations placed stringent limits on the effective number of
relativistic species $N_{\rm eff}^{\rm CMB}$. The parameter
$N_{\rm eff}^{\rm CMB}$ is an estimate of
the total energy density
contained in relativistic particles at recombination, parametrized in terms of the number
of effective two-component neutrinos $N_\nu$. The analysis done by the PLANCK collaboration reveals \cite{Ade:2015xua}
\begin{equation}
\label{CMBlimit}
N_{\rm eff}^{\rm CMB} = 3.15 \pm 0.23,
\end{equation}
where this value and its best-fit estimate are inferred in a Bayesian treatment
by combining the Planck measurements with other cosmological data sources. While $N_{\rm eff}^{\rm CMB}$ includes any particles that are relativistic at recombination, we focus on the case that such ``dark radiation" consists entirely of neutrinos and leave more exotic possibilities to future work.

Though not necessarily the case, it is often assumed that CMB measurements of $N_{\rm eff}^{\rm CMB}$ provide a direct probe of the number of neutrinos $N_\nu$. CMB observations are generally only sensitive to the total
neutrino energy density at recombination through its effect on the expansion rate. The neutrino
energy density at recombination is therefore the direct CMB observable, which does not depend only on $N_\nu$, but also on the neutrino temperature $T_\nu$.
The equivalence between $N_\nu$ and the value of $N_{\rm eff}^{\rm CMB}$ derived
from CMB observations assumes a standard neutrino thermal history, which we refer to henceforth as the ``standard model" (SM) temperature of neutrinos $T_{\nu \rm SM}$.
The equivalence between $N_\nu$ and $N_{\rm eff}^{\rm CMB}$ is broken if nonstandard processes take place after neutrino decoupling,
resulting in a temperature deviation from the usual $T_{\nu \rm SM}$. 

The possibility that the neutrinos might have a nonstandard temperature has a rich history and has been explored
in numerous papers \cite{Kolb:1986nf, Serpico:2004nm, Ho:2012ug, Ho:2012br, Boehm:2013jpa, Steigman:2013yua, Nollett:2013pwa, Nollett:2014lwa}.
As an example of such a scenario, particles that decay after the neutrinos decouple at a temperature of a few MeV
may raise the photon temperature relative to the neutrinos, giving an effective neutrino temperature that
is lower than that of the standard cosmological scenario.  A similar effect can also occur for MeV-mass
dark matter, which can stay in thermal equilibrium long enough to heat the photons relative to the neutrinos. 

In this more general case, the statement that $N_{\rm eff}^{\rm CMB} = N_\nu$ no longer holds true and is now generalized to
\begin{equation}\label{relation:cmb-bbn}
N_{\rm eff}^{\rm CMB} T_{\nu \rm SM}^4 = N_\nu T_\nu^4,
\end{equation}
where the total neutrino energy density (as long as the neutrinos are relativistic) is proportional to $N_\nu T_\nu^4$. The usual case of $N_{\rm eff}^{\rm CMB} = N_\nu$ is recovered if a standard-model
neutrino temperature $T_{\nu SM}$ is assumed, with the more general treatment given by Eq. (\ref{relation:cmb-bbn}). 

This degeneracy between $N_{\rm eff}^{\rm CMB}$ and $N_\nu$ can however be disentangled by combining CMB and BBN observables. For example, \citet{Nollett:2013pwa} recently used the combination of BBN abundance predictions and CMB measurements to constrain electromagnetically coupled dark matter
particles that raise the photon temperature relative to that of the neutrinos. They also showed in \cite{Nollett:2014lwa} that the opposite effect can be accomplished if a coupling is introduced between the dark matter sector(s) and neutrinos. In this paper, we consider the most general case, in which $N_\nu$ and $T_\nu$ are treated as free parameters, and then
determine the observational constraints obtained from a combination of BBN and the CMB.  

Before the neutrinos decouple
at the temperature $T_d \approx 2-3$ MeV, the weak
rates ensure that $T_\nu = T_\gamma$.  We make the assumption that the change in $T_\nu$ is induced after $T_d$, but
before BBN begins at $T_\gamma \sim 1$ MeV, and that the neutrino temperature subsequently evolves in the standard
way as the inverse of the scale factor.  This is admittedly a narrow window over which the change is assumed to occur,
and it is also true that the onset of BBN is not a sudden process.  We discuss these issues in more detail below.

The paper is organized as follows: in section \ref{sec:cosmo_params} we discuss how $T_\nu$ and $N_\nu$ affect BBN and CMB observables, while in section \ref{sec:numerical} we explore these effects more 
precisely using numerical simulations of BBN and its effects on primordial abundance predictions of $^4$He and deuterium. 
Combining these results with observational limits on the primordial element abundances, we derive corresponding limits on
$T_\nu$ and $N_\nu$ and then examine the effects on $N_{\rm eff}^{\rm CMB}$. We find that even current observational bounds on $^4$He and deuterium, when combined with CMB limits, can be consistent with
one extra sterile neutrino for a change in $T_\nu$ from the standard model value of only $\sim -5\%$, while
two additional sterile neutrinos are ruled out. We discuss certain implications of this analysis and conclude in section \ref{sec:conclusions}.

\section{The effects of $N_\nu$, $N_{\rm eff}^{\rm CMB}$ and $T_\nu$ on the CMB and BBN }\label{sec:cosmo_params}

Currently, the number of neutrinos $N_\nu$ can be probed in two separate
eras of cosmic evolution, both producing distinct and independent observables. These
two eras are $(i)$ {\it Big Bang nucleosynthesis} (BBN), when light nuclear
elements are produced and $(ii)$ {\it Recombination},
when electrically neutral atoms first form allowing photons to free-stream and produce the cosmic microwave background (CMB) radiation. These two eras differ by orders of magnitude in energy, with BBN occurring around $\sim 1$ MeV and CMB decoupling occurring at the eV scale.

Consider first the standard neutrino thermal history.  At sufficiently high temperatures, the weak interactions
are sufficiently rapid to keep the neutrinos in equilibrium with the thermal background, so that $T_\nu = T_\gamma$.
At a temperature $T_d \approx 2-3$ MeV, the neutrinos decouple from the thermal background.
Then $T_\nu$
is solely affected by the expansion of the universe, scaling as $T_\nu \propto a^{-1}$, where
$a$ is the cosmic scale factor.

When the temperature drops below the mass of the electron, $e^+ e^-$ pairs annihilate, heating
the photons relative to the neutrinos, and producing a final ratio of
\begin{equation}
\label{Tnustandard}
T_{\nu \rm SM} = (4/11)^{1/3} T_\gamma.
\end{equation}
The assumption made in this scenario is that there are
no processes that modify either the neutrino temperature or
the photon temperature between neutrino decoupling and BBN.
We refer to this evolution as the standard model evolution, hence our use of the SM subscript.

Note that the neutrinos are partially heated by the $e^+ e^-$ annihilations.  Neither
the process of neutrino decoupling, nor the annihilations of the $e^+e^-$ occurs sharply
at a single temperature, and the resulting overlap between these two processes \cite{Steigman:2013yua} results
in
\begin{equation}
\label{Tnuheat}
T_{\nu \rm SM} > (4/11)^{1/3} T_\gamma.
\end{equation}
This effect is usually
absorbed into the definition of $N_\nu$ rather than $T_\nu$, giving an effective neutrino number
of $N_{\rm eff} = 3.046$ \cite{Dolgov:2002wy,Mangano:2005cc}.  (See also
the more recent discussion by de Salas and Pastor \cite{deSalas:2016ztq},
which gives a similar value of $N_{\rm eff} = 3.045$).  However, this effect is very small compared to the large changes in $N_\nu$
and $T_\nu$ considered here, so we ignore it in what follows.

Now suppose that this neutrino thermal history is modified by some process similar to those discussed in Refs.
\cite{Kolb:1986nf, Serpico:2004nm, Ho:2012ug, Ho:2012br, Boehm:2013jpa, Steigman:2013yua, Nollett:2013pwa, Nollett:2014lwa}.
Rather than specifying a particular process, we will attempt to keep the discussion as general as possible.
Note that the measurable quantity that affects both BBN and the CMB is not the absolute
value of $T_\nu$, but the ratio of $T_\nu$ to $T_\gamma$, since all calculations for BBN and the CMB
are scaled off of the background photon temperature.  For simplicity, we treat any change in $T_\nu/T_\gamma$
as an effective change in $T_\nu$.  However, note that $T_\nu/T_\gamma$ can be altered by
changing {\it either} $T_\nu$ or $T_\gamma$.  We assume here that a physical process occurring after $T_d$,
but before the beginning of BBN, alters $T_\nu/T_\gamma$ by a fixed amount.  This can be accounted for by
changing the neutrino temperature from its standard model value, given by Eq. (\ref{Tnuheat}), to a new
value, which we treat as a free parameter in the calculations.  Then $T_\nu$ evolves in the standard way (inversely
proportional to the scale factor), but in such a way that
\begin{equation}
\label{stepT}
T_\nu/T_{\nu SM} = constant \ne 1,
\end{equation}
from the beginning
of BBN up to the present day.

A second possibility is that $T_\nu$ takes its standard model value
during BBN, but then changes between BBN and decoupling.  For example, entropy release between BBN and decoupling
results in an increase in $T_\gamma$, so an effective decrease in $T_\nu$ at a given value of $T_\gamma$.
These scenarios are straightforward to analyze:  BBN proceeds in the standard way, but the CMB limits are altered by changing
the value of $N_{\rm eff}^{\rm CMB}$ relative to $N_\nu$ as given by Eq. (\ref{NeffCMB}) below.  We will not examine
such scenarios in detail here.

The model we are examining is admittedly limited in applicability, since the window between $T_d$ and the onset of BBN is narrow.
However, there are certainly physical processes which can yield the case we investigate here in appropriate limits.  For instance,
a decaying particle with a lifetime much shorter than the age of the universe at $T_d$ will heat the photons relative to the neutrinos
in the exponential tail of its decay, with negligible effects at later times when BBN begins.  An electromagnetically-coupled
WIMP with a mass of $1-10$ MeV would annihilate largely after neutrino decoupling and before the onset of BBN, as noted
in Ref. \cite{Nollett:2013pwa}.  Of course, the most general
possible case would allow for $T_\nu/T_{\nu SM}$ to evolve before, during, and after BBN, but such a general treatment is
beyond of the scope of this paper.

The CMB measurements are sensitive to the total energy density at the epoch of recombination; the neutrino
energy density enters into this calculation only through its total energy density, which is simply proportional
to $N_\nu T_\nu^4$.  If we allow both $N_\nu$ and $T_\nu$ to have values that differ from their standard-model
values, then the number of neutrinos inferred from CMB measurements will be
\begin{equation}
\label{NeffCMB}
N_{\rm eff}^{\rm CMB} = N_\nu (T_{\nu}/T_{\nu SM})^4.
\end{equation}
We can therefore quantify a possible shift in the neutrino temperature through the ratio of $T_{\nu}/T_{\nu SM}$.

The processes involved in Big Bang Nucleosynthesis also depend on both $N_\nu$ and $T_\nu$, but in a more complex way than the CMB observables do.
(For a recent review of BBN, see Ref. \cite{Cyburt:2015mya}).
First, the element abundances depend on the expansion rate during BBN given by
\begin{equation}
H^2 = \frac{8 \pi G}{3}(\rho_\gamma + \rho_{e^+e^-} + \rho_{\nu \bar \nu}),
\end{equation}
where the three terms that contribute to the total energy density are those of the photons,
electron-positron pairs and neutrinos, respectively. The neutrino term includes the contributions of all three standard-model neutrinos as well as any possible nonstandard (e.g., sterile)
neutrinos. If the expansion rate was the only place
where the neutrino temperature entered into the BBN calculations,
then the primordial element abundances would only depend on the total neutrino energy density, just like the CMB
observations. That this is not the case is the fundamental reason we can break the degeneracy between $N_{\rm eff}^{\rm CMB}$ and $N_\nu$.

Beyond its role in the expansion rate during BBN, the neutrino temperature also plays a crucial role in the weak
interaction rates, which determine the light element abundances.  Down to temperatures of $\sim 1$ MeV, the protons and neutrons are kept in thermal equilibrium via
the following weak interaction
processes
\begin{eqnarray}\label{weak-interactions}
n+\nu _{e} &\leftrightarrow &p+e^{-},  \notag \\
n+e^{+} &\leftrightarrow &p+\bar{\nu}_{e},  \notag \\
n &\leftrightarrow &p+e^{-}+\bar{\nu}_{e}.
\end{eqnarray}
The total rates for the conversion of neutrons to protons and protons to neutrons are
\begin{eqnarray}
\label{n->p}
&&\lambda_{n\rightarrow p}
= A \int_{m_e}^\infty dE_e
\frac{E_e|p_e|}{1+\exp[E_e/kT_e]} \nonumber \\
&&\times \left\{ \frac{(E_e+Q)^2}{1+\exp[-(E_e+Q)/kT_{\nu_e}]}+
\frac{(E_e-Q)^2\exp(E_e/kT_e)}{1+ \exp[(E_e-Q)/kT_{\nu_e}]}
\right\},
\end{eqnarray}
and
\begin{equation}
\lambda_{p\rightarrow n} =  \lambda_{n\rightarrow p}(-Q),
\end{equation}
respectively,
where $Q = m_n - m_p$, and
the subscripts $e$ and $\nu_e$ denote the quantities associated with the electron
and the electron-neutrino, respectively.
The constant $A$ is determined from the requirement that
$\lambda_{n \rightarrow p}(T,T_{\nu_e} \rightarrow 0) = 1/\tau_n$ (the neutron decay
rate).

The important point is that
these weak rates are sensitive
to the neutrino temperature $T_{\nu }$, where we assume that neutrino mixing gives all three neutrinos the same temperature.
(The actual effect on the various element abundances is discussed in detail in the next section).
This dependance on $T_\nu$ allows a combination of BBN abundance predictions and CMB observations
to yield complementary limits on $T_\nu$ and $N_\nu$ when these quantities are varied independently.

The two observables we discuss here (BBN and CMB) are not the only means by which the degeneracy between $N_\nu$ and
$T_\nu$ can be broken.  A third possibility involves large-scale structure constraints on the neutrino mass.  Since these
observations are based on an epoch at which the neutrinos have become nonrelativistic, they are actually sensitive
to the quantity $T_{\nu}^3 \Sigma_\nu m_{\nu}$, where the sum is over all three types of neutrinos, again assumed to have
a single common temperature $T_\nu$.  A discussion of these limits and their relation to CMB and BBN limits is beyond the
scope of this paper, but see the analysis in Ref. \cite{Ho:2012br}.

\section{Numerical results, observational bounds, and combined BBN/CMB constraints}\label{sec:numerical}

In order to investigate the interplay between a varying
neutrino number $N_\nu$ and neutrino temperature $T_\nu$ on light element abundances from BBN,
we solve the rate equations and cosmological evolution numerically and present our results in this section.
We used the computer code {\it AlterBBN} \cite{Arbey:2011nf} originally written by A. Arbey,
and later modified by K. P. Hickerson \cite{Hickerson:2016BBNCODE}. Our version, modified from Hickerson's version 1.6, along with a supplementary explanation of our analysis is available at \cite{Galvez:2016BBNCODE}.

In our simulations, we allow $N_\nu$ and $T_\nu$ to vary separately. We take $T_\nu$ to be the same for all three standard-model neutrinos and any additional
sterile neutrinos, which will be the case as long as there is sufficient mixing between all of the neutrino sectors. This corresponds, for instance, to the most interesting cases of mixing with sterile neutrinos to provide a possible explanation for the tension in short-baseline neutrino
oscillation experiments (see, e.g., Ref. \cite{Ho:2012br} and references therein).

In our simulations we take a baryon to photon ratio of $\eta = 6.19 \times 10^{-10}$ \cite{Agashe:2014kda} and a neutron lifetime of $\tau_n = 880.3$ seconds \cite{Agashe:2014kda}, 
and derive the primordial element abundances as a function of two parameters: $N_\nu$, and
a nonstandard neutrino temperature $T_\nu$.
We parametrize the shift of the neutrino temperature 
relative to the standard-model neutrino temperature as the ratio $T_\nu /T_{\nu \rm SM}$.
In order to keep the results as general as possible,
we do not assume a particular model or mechanism for
the nonstandard value of $T_\nu$; instead, as noted in the previous section, we assume that the neutrino
temperature differs by a constant ratio from the standard model neutrino temperature throughout BBN, and that this
same ratio is maintained up to the present.
Aside from this fixed choice of $T_\nu/T_{\nu SM} \ne 1$, we assume
that the neutrino temperature obeys the standard evolution throughout and after BBN,
i.e., it decreases as the inverse of the scale factor. One may consider other
deviations from our scenario for $T_\nu$; however, we leave such possibilities for future work and focus on the scenario listed above.

Since we assume three standard neutrinos plus an undefined additional contribution to $N_\nu$, it is
reasonable to take
$N_\nu \ge 3$ to be a physical lower bound.
Note, however, that there are brane-world scenarios which
achieve a negative change in the relativistic energy density \cite{Bratt:2002xt}, so
for completeness we allow $N_\nu$ to vary in the range
$2 \le N_\nu \le 5$.

\begin{figure}
    \centering
	\includegraphics[scale=0.6]{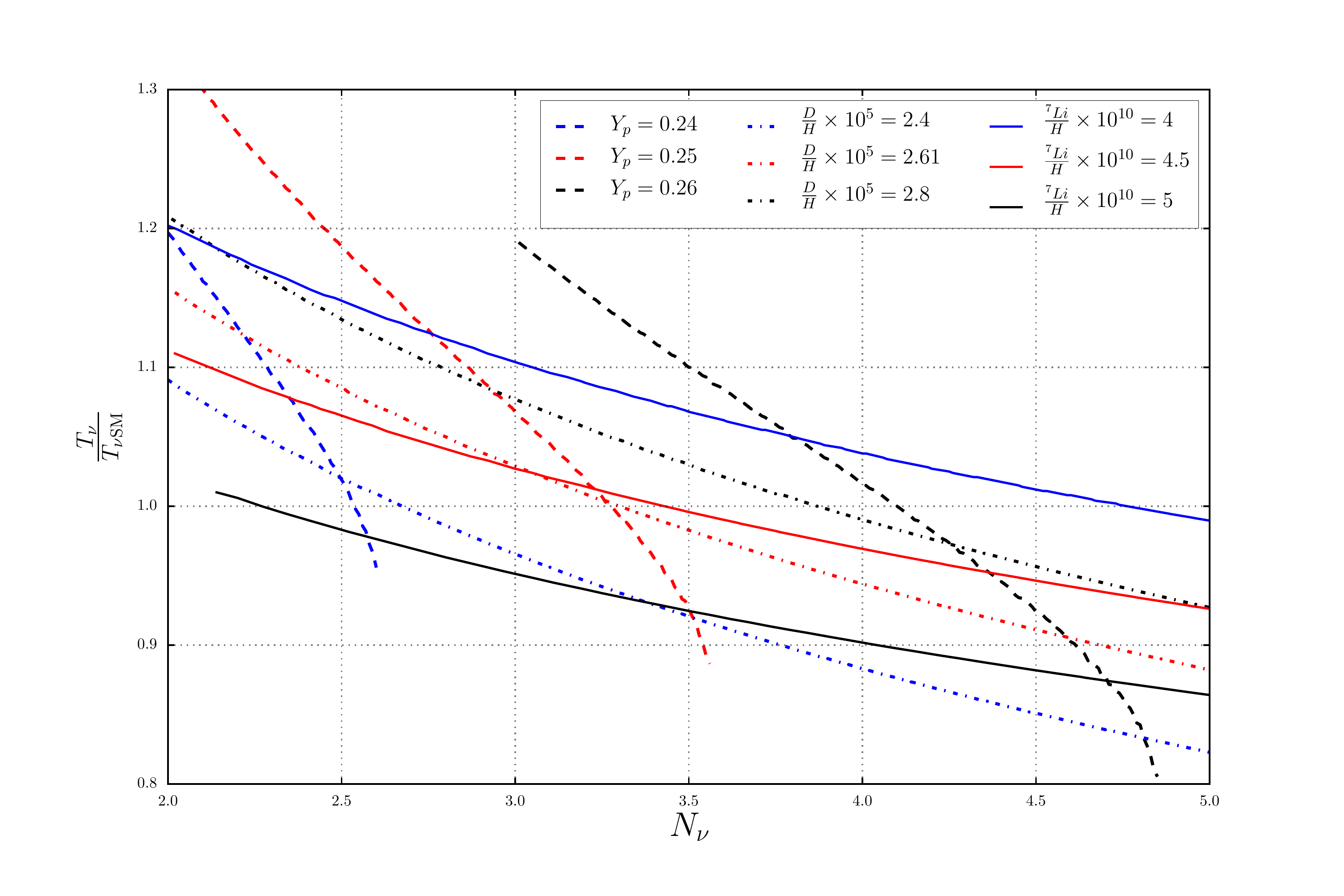}
    \caption{Predicted primordial abundances of $^4$He, deuterium, and $^7$Li in the plane defined by the neutrino number, $N_\nu$,
    and the ratio of the neutrino temperature to its standard-model value, $T_\nu/T_{\nu SM}$.}
    \label{fig1}
\end{figure}

 \begin{figure}
    \centering
	\includegraphics[scale=0.7]{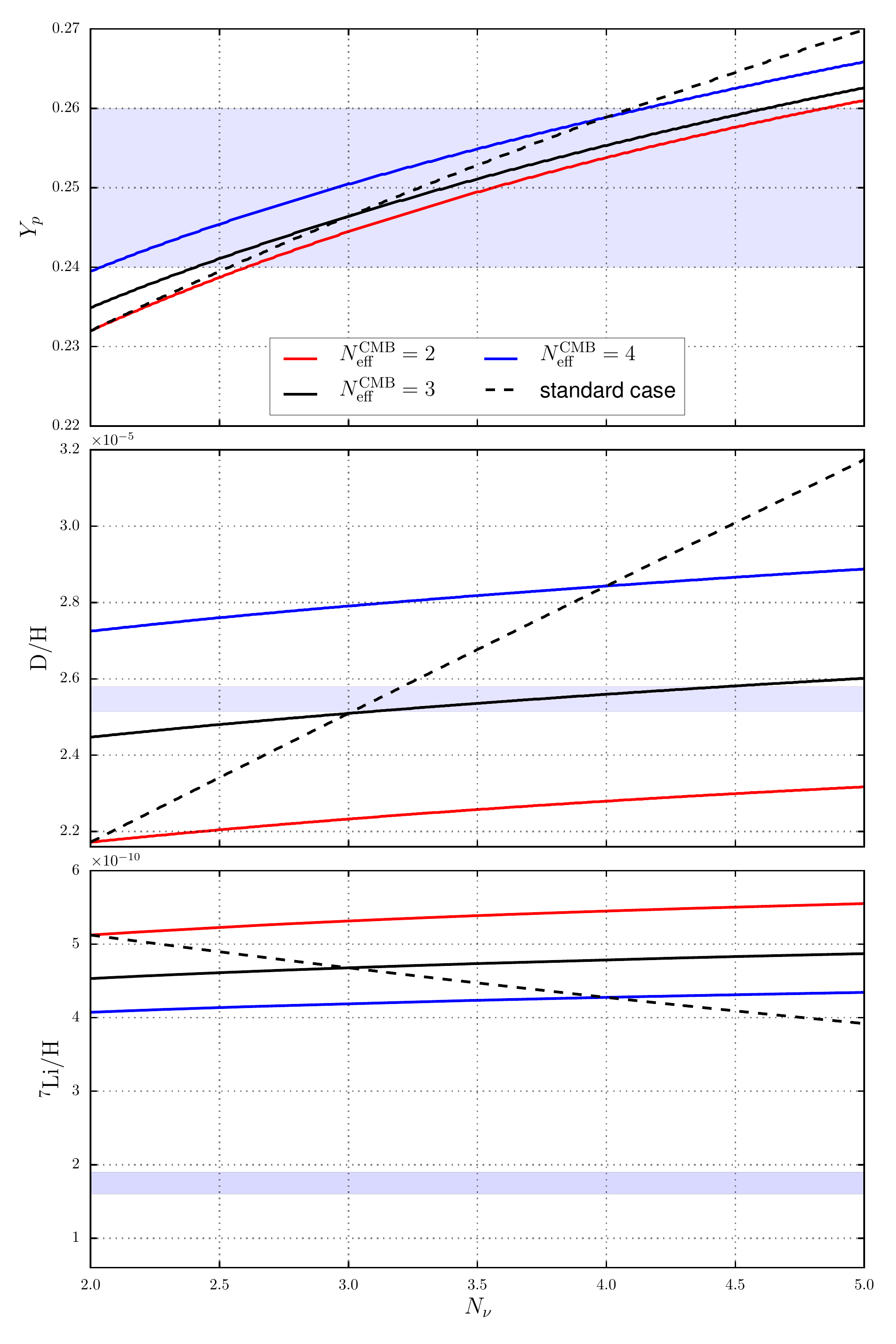}
    \caption{Predicted primordial abundances
    of $^4$He, deuterium, and $^7$Li
    as a function of the number of relativistic neutrinos $N_\nu$
    when the neutrino temperature is allowed to vary from its standard-model
    value.  The neutrino temperature is parametrized in
   terms of $N_{\rm eff}^{\rm CMB}$ as defined in Eq. (\ref{NeffCMB}); solid
   curves give abundances for the indicated values of $N_{\rm eff}^{\rm CMB}$.
    Dashed curve gives abundances for the standard-model neutrino temperature.
    Shaded regions correspond to the observational abundance limits quoted in Eqs. (\ref{Dlimit}) - (\ref{Lilimit}).}
    \label{fig2}
\end{figure}

In Fig. 1, we show the primordial $^4$He mass fraction, $Y_p$, and the deuterium and $^7$Li number densities
relative to hydrogen, as a function of $N_\nu$ and $T_\nu/T_{\nu SM}$.  For the standard model
value of the neutrino temperature, $T_\nu/T_{\nu SM}= 1$, we obtain the familiar result that $Y_p$ increases with $N_\nu$, because the increased
expansion rate causes the weak rates to freeze out at a higher temperature, resulting in more neutrons, and nearly all of these
neutrons (modulo free neutron decay) end up bound into $^4$He.  The deuterium abundance
also increases with $N_\nu$ at fixed $T_\nu$, as the increased expansion rate allows less time for the deuterium to fuse
into heavier elements.  On the other hand, the $^7$Li abundance decreases with increasing
$N_\nu$.

The effect of altering $T_\nu$ at fixed $N_\nu$ is not as obvious, because an increase in $T_\nu$ results in both
an increase in the weak rates and an increase in the expansion rate.  From
Fig. 1, we see that increasing $T_\nu$ at fixed $N_\nu$ results in a net increase in $Y_p$, indicating that the effect
of increasing the expansion rate (which increases $Y_p$) dominates the effect of increasing the weak rates (which decreases
$Y_p$).  Note, however that these two effects begin to cancel for $T_\nu/T_{\nu SM} < 0.9$, at which point $Y_p$ becomes
less sensitive to $T_\nu/T_{\nu SM}$.  The behavior of deuterium and $^7$Li is much more straightforward,
since these two nuclides are primarily sensitive to the overall expansion rate.  Thus, $D/H$ increases with the increased
expansion rate produced by an increased value of $T_\nu/T_{\nu SM}$, while $^7$Li decreases.

These results are not directly comparable to previous studies, but a subset of our results is in qualitative agreement
with those of Nollett and Steigman \cite{Nollett:2013pwa} for electromagnetically-coupled WIMPS.
In Ref. \cite{Nollett:2013pwa}, WIMP annihilation heats the photons relative to the neutrinos, producing a net
decrease in $T_\nu/T_\gamma$.  Our case with $N_\nu = 3$ and $T_\nu/T_{\nu SM} < 1$ corresponds roughly
to the effect of a 1-10 MeV WIMP, designated
``region II" in Fig. 4 of Ref. \cite{Nollett:2013pwa}.  In this portion of the parameter space,
Nollett and Steigman find a decrease in the production of deuterium and $^4$He and an increase in $^7$Li,
in agreement with the behavior we see in Fig. 1 as $T_\nu/T_{\nu SM}$ decreases.

Note that while $T_\nu/T_{\nu SM}$ and $N_\nu$ are the two parameters that enter
directly into the BBN calculation, they are not the most useful to use in our analysis.
Instead, we take $N_{\rm eff}^{\rm CMB}$ to be one of our parameters, as this is the effective
number of neutrinos measured by the CMB.  Eq. (\ref{NeffCMB}) then leaves only one free parameter,
which we can take to be either the neutrino number or temperature.  Since it is the neutrino number
which is the physically relevant quantity, we adopt it as our second parameter.
For a given set of values of $N_\nu$ and $N_{\rm eff}^{\rm CMB}$, the value
of $T_\nu/T_{\nu SM}$ can be determined from Eq. (\ref{NeffCMB}).

Now consider the observational limits on the
primordial $^4$He mass fraction, $Y_p$, and the deuterium
to hydrogen ratio $D/H$.
For deuterium, the Particle Data group gives \cite{Agashe:2014kda} $D/H = (2.53 \pm 0.04) \times 10^{-5}$, while
a more recent measurement by Cooke et al. gives \cite{Cooke:2016rky} $D/H = 2.547 \pm 0.033 \times 10^{-5}$.
We will therefore take our limit to be
\begin{equation}
\label{Dlimit}
D/H = (2.55 \pm 0.03) \times 10^{-5}.
\end{equation}

Recent estimates of the primordial $^4$He mass fraction include those of Izotov et al. \cite {Izotov:2014fga},
$Y_p = 0.2551 \pm 0.0022$, and Aver et al. \cite{Aver:2015iza}, $Y_p = 0.2449 \pm 0.0040$, while the Particle Data
Group limit is \cite{Agashe:2014kda} $Y_p = 0.2465 \pm 0.0097$. Given the discrepancy between these numbers, we will adopt
as our limit
\begin{equation}
\label{Helimit}
Y_p = 0.25 \pm 0.01.
\end{equation}

It is well-known that the current BBN predictions for the primordial $^7$Li abundance differ
significantly from the
observationally-inferred values, with the BBN predictions a factor of 3
or more above the observed values.  This has been dubbed the ``lithium problem" (see, e.g., Ref.
\cite{Fields:2011zzb} 
for a recent review).  Thus, we will not use $^7$Li to constrain the models examined here; instead, our main
interest will be to determine whether these models can ameliorate the lithium problem.  To that end, we will adopt
a $^7$Li abundance of \cite{Agashe:2014kda}
\begin{equation}
\label{Lilimit}
Li/H = (1.6 \pm 0.3) \times 10^{-10}.
\end{equation}

In Fig. 2, we present
the predicted element abundances as a function of $N_\nu$, for a variety of $N_{\rm eff}^{\rm CMB}$ values (solid curves).
The dashed curve corresponds to the standard temperature case,
$T_\nu /T_{\nu \rm SM} = 1$.  As is clear in the figure, this curve intersects each $N_{\rm eff}^{\rm CMB}$ curve at the point
$N_{\rm eff}^{\rm CMB} = N_\nu$. 
Curves of constant $N_{\rm eff}^{\rm CMB}$, as
defined in Eq. (\ref{NeffCMB}), correspond to curves of constant neutrino energy density. Thus, tracing the element
abundances
along each solid curve allows us to see the effect of changing both the neutrino temperature and number in such a way that
the neutrino energy density is unchanged.
In this case, the only effect on the primordial element abundances comes from the change in the weak rates.
Decreasing $N_\nu$ at fixed $N_{\rm eff}^{\rm CMB}$ corresponds to increasing $T_\nu$.  In the case of $^4$He, for example, this
results in a decrease in ${Y_p}$,
since the increased neutrino temperature increases
the weak rates, allowing them to stay in thermal equilibrium longer and reducing the final neutron abundance. 

Now consider the observational bounds from BBN.  When
we move beyond the standard case for the neutrino
temperature (dashed curves in Fig. 2) to allow both $N_\nu$ and $N_{\rm eff}^{\rm CMB}$ to vary independently (solid curves), we see the
largest effect is a relaxation of the bounds from deuterium.  The standard case corresponds to a narrow window near $N_\nu = 3$, while the model
we consider here allows
$N_\nu$ to have any value in the range we have investigated ($2 \le N_\nu \le 5$).
This result is easy to understand; for the model discussed here, an increase an $N_\nu$ can be compensated by a decrease in $T_\nu$ to
give an unchanged expansion rate.  Since deuterium depends almost entirely on the expansion rate, with only a very weak dependence
on the weak rates, we can always find a value of $N_{\rm eff}^{\rm CMB}$ for any given $N_\nu$ to give a deuterium abundance
in the desired range.

Since $Y_p$ is strongly dependent on both the expansion rate and the weak rates, compensating
a change in $N_\nu$ with a change in $T_\nu$ to leave the expansion rate fixed does not leave the $^4$He abundance unchanged.
Consequently, primordial helium continues to provide an upper bound on $N_\nu$ even with the freedom to alter the neutrino temperature.
Varying the neutrino temperature relaxes this bound somewhat,
but even for $N_{\rm eff}^{\rm CMB}$ as low as 2, we still have the upper
bound $N_\nu < 5$.

\begin{figure}
    \centering
    \includegraphics[scale=0.6]{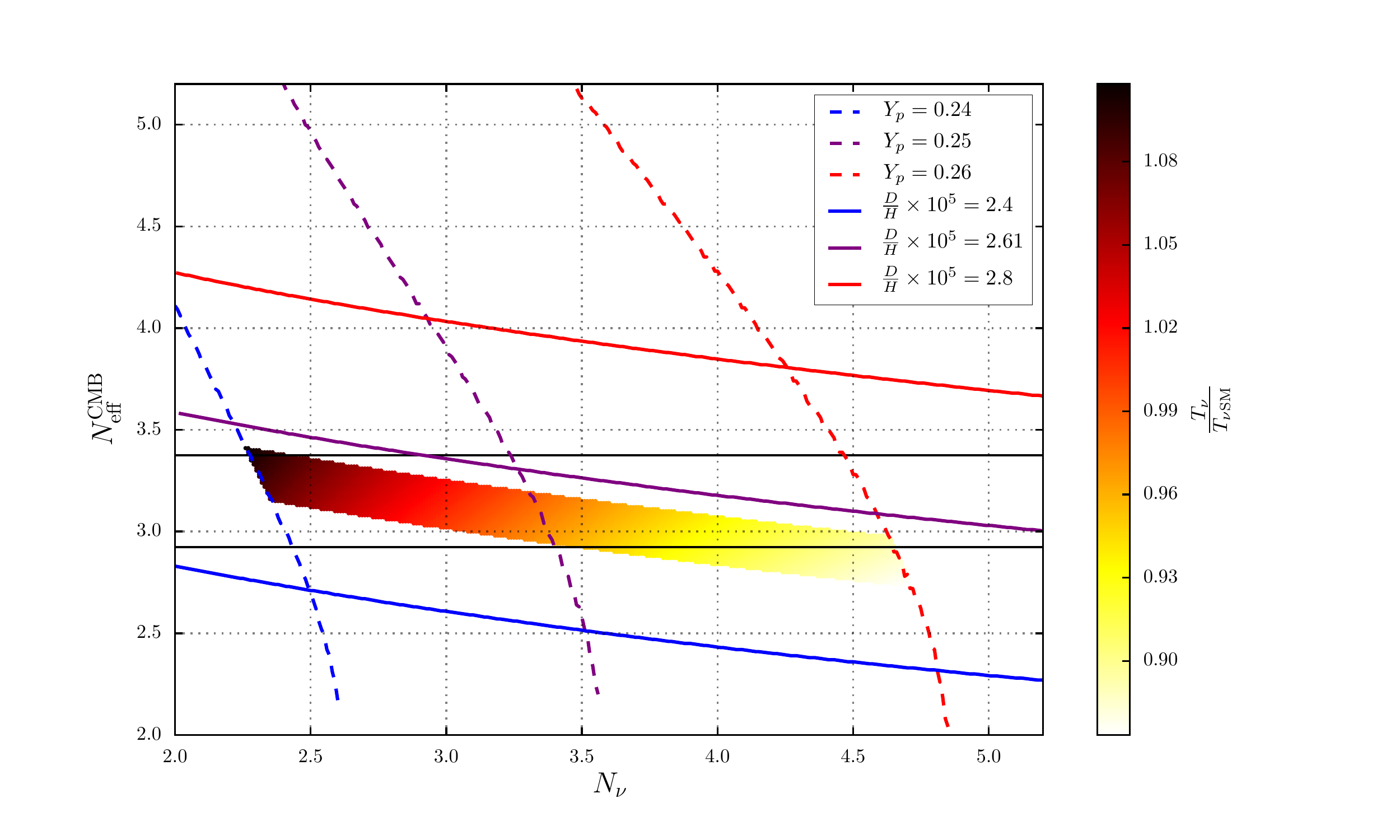}
    \caption{Allowed region in the $N_\nu$, $N_{\rm eff}^{\rm CMB}$ plane given by combined observational
limits on deuterium and $^4$He, along with CMB bounds on $N_{\rm eff}^{\rm CMB}$.  Red-orange-yellow quadrilateral is allowed by the BBN
bounds on deuterium and $^4$He (Eqs. \ref{Dlimit} - \ref{Helimit}).  Black horizontal lines give upper and lower bounds on $N_{\rm eff}^{\rm CMB}$ from CMB
observations (Eq. \ref{CMBlimit}).  The overlap between these two regions is allowed
by both BBN and CMB.  Dashed and solid curves correspond
to the indicated $^4$He and deuterium abundances, respectively, while the heat map gives the value of $T_\nu/T_{\nu SM}$ corresponding to
a given region of the parameter space allowed by BBN.}
\label{fig3}
\end{figure}

In Fig. 3 we combine the deuterium and $^4$He limits to derive overall
constraints in the $N_\nu$, $N_{\rm eff}^{\rm CMB}$ plane.  This figure shows the complementarity of
these two sets of limits, with deuterium giving the upper and lower bounds on $N_{\rm eff}^{\rm CMB}$, and $Y_p$ giving the
upper and lower bounds on $N_\nu$. Adding the CMB bound from Eq. (\ref{CMBlimit})
tightens the lower bound on $N_{\rm eff}^{\rm CMB}$, but otherwise has very little effect on the excluded region.
In particular, even when we include all three sets of limits (deuterium, $^4$He, and the CMB),
the value of $N_\nu$ can be as large as 4.5, thus allowing on additional
sterile neutrino.  On the other hand, a value
of $N_\nu = 5$ (two additional sterile neutrinos) is ruled out.
Fig. 3 illustrates the additional constraining power of BBN beyond what is available with the
CMB alone.  When $T_\nu$ is allowed to vary freely, a given value of $N_{\rm eff}^{\rm CMB}$ from the CMB no longer provides
a constraint on $N_\nu$.  But adding the BBN constraint reestablishes the
upper bound on $N_\nu$; as we have already noted, this upper bound
is derived primarily from limits on $^4$He, rather than deuterium.

In Fig. 3 we have also superimposed a heat map
to illustrate the value of $T_\nu/T_{\nu SM}$ corresponding to
a given region of the allowed parameter space.  The region for which
$N_\nu > 3$ corresponds to a value of $T_\nu$ smaller than its standard-model value.
Interesting effects are achieved with a very small change in $T_\nu$.  For instance, the point
corresponding to $N_{\rm eff}^{\rm CMB} = 3$ and $N_\nu = 4$ corresponds to a $\sim 5\%$
decrease in $T_\nu$ relative to its standard model value.

We can also use Fig. 3 to derive the combined BBN-CMB {\it lower} bound on $N_\nu$ when the neutrino
number and temperature are allowed to vary independently: $N_\nu > 2.3$.  This limit is considerably
less interesting than our upper bound, as we already have $N_\nu \ge 3$ from the observed
neutrinos.  However, as noted earlier, this does serve as a constraint on nonstandard models such as those
discussed in Ref. \cite{Bratt:2002xt}.

It is interesting to see whether the joint variation of $N_\nu$
and $T_\nu$ can ameloriate or solve the lithium problem.  It is clear from Fig. 2 that although some combinations
of $N_\nu$ and $N_{\rm eff}^{\rm CMB}$ can reduce the predicted $^7$Li abundance, this reduction is short of
what is needed to close the gap between prediction and observation.  Furthermore, the largest reductions
in the predicted abundance lie in regions of parameter space that are excluded by the deuterium
and $^4$He observations.

\section{Conclusions}\label{sec:conclusions}

While observations of the CMB yield very precise limits
on cosmological parameters, our results show that Big Bang nucleosynthesis remains
an indispensible tool.  For models in which the neutrino number and temperature can both
vary, the CMB alone cannot produce any limits on $N_\nu$, while a combination of the CMB and BBN
yields a very useful bound.

In the models examined here, a value of the neutrino number as determined from the CMB of
$N_{\rm eff}^{\rm CMB} \approx 3$ can be consistent with a true neutrino number, $N_\nu$, as large as 4, thus
allowing for an additional sterile neutrino.  Such a model requires a reduction in the neutrino temperature
of approximately 5\% relative to the standard model neutrino temperature.
However, a value of $N_\nu =5$ is ruled out for any value of $N_{\rm eff}^{\rm CMB}$.

The obvious direction for future investigation would involve more complex behavior for the evolution
of the neutrino temperature, both during and following BBN.  Some of these types of behavior have
been discussed previously in
Refs. \cite{Kolb:1986nf, Serpico:2004nm, Ho:2012ug, Ho:2012br, Boehm:2013jpa, Steigman:2013yua, Nollett:2013pwa,
Nollett:2014lwa},
but these studies do not by any means exhaust all of the interesting possibilities.

 \section*{Acknowledgements}
R.G. is grateful to Celine Boehm, Evan Grohs, Jeff McMahon and Aaron Pierce for useful comments. R.G. was supported in part by the National Science Foundation (AST-1358862).
R.J.S. was supported in part by the Department of Energy
(DE-FG05-85ER40226).

\bibliography{references}

\begin{thebibliography}{24}
\expandafter\ifx\csname natexlab\endcsname\relax\def\natexlab#1{#1}\fi
\expandafter\ifx\csname bibnamefont\endcsname\relax
  \def\bibnamefont#1{#1}\fi
\expandafter\ifx\csname bibfnamefont\endcsname\relax
  \def\bibfnamefont#1{#1}\fi
\expandafter\ifx\csname citenamefont\endcsname\relax
  \def\citenamefont#1{#1}\fi
\expandafter\ifx\csname url\endcsname\relax
  \def\url#1{\texttt{#1}}\fi
\expandafter\ifx\csname urlprefix\endcsname\relax\def\urlprefix{URL }\fi
\providecommand{\bibinfo}[2]{#2}
\providecommand{\eprint}[2][]{\url{#2}}

\bibitem[{\citenamefont{Steigman et~al.}(1977)\citenamefont{Steigman, Schramm,
  and Gunn}}]{Steigman:1977kc}
\bibinfo{author}{\bibfnamefont{G.}~\bibnamefont{Steigman}},
  \bibinfo{author}{\bibfnamefont{D.~N.} \bibnamefont{Schramm}},
  \bibnamefont{and} \bibinfo{author}{\bibfnamefont{J.~E.} \bibnamefont{Gunn}},
  \bibinfo{journal}{Phys. Lett.} \textbf{\bibinfo{volume}{B66}},
  \bibinfo{pages}{202} (\bibinfo{year}{1977}).

\bibitem[{\citenamefont{Hinshaw et~al.}(2013)}]{Hinshaw:2012aka}
\bibinfo{author}{\bibfnamefont{G.}~\bibnamefont{Hinshaw}} \bibnamefont{et~al.}
  (\bibinfo{collaboration}{WMAP}), \bibinfo{journal}{Astrophys. J. Suppl.}
  \textbf{\bibinfo{volume}{208}}, \bibinfo{pages}{19} (\bibinfo{year}{2013}),
  \eprint{1212.5226}.

\bibitem[{\citenamefont{Ade et~al.}(2015)}]{Ade:2015xua}
\bibinfo{author}{\bibfnamefont{P.~A.~R.} \bibnamefont{Ade}}
  \bibnamefont{et~al.} (\bibinfo{collaboration}{Planck})
  (\bibinfo{year}{2015}), \eprint{1502.01589}.

\bibitem[{\citenamefont{Kolb et~al.}(1986)\citenamefont{Kolb, Turner, and
  Walker}}]{Kolb:1986nf}
\bibinfo{author}{\bibfnamefont{E.~W.} \bibnamefont{Kolb}},
  \bibinfo{author}{\bibfnamefont{M.~S.} \bibnamefont{Turner}},
  \bibnamefont{and} \bibinfo{author}{\bibfnamefont{T.~P.}
  \bibnamefont{Walker}}, \bibinfo{journal}{Phys. Rev.}
  \textbf{\bibinfo{volume}{D34}}, \bibinfo{pages}{2197} (\bibinfo{year}{1986}).

\bibitem[{\citenamefont{Serpico and Raffelt}(2004)}]{Serpico:2004nm}
\bibinfo{author}{\bibfnamefont{P.~D.} \bibnamefont{Serpico}} \bibnamefont{and}
  \bibinfo{author}{\bibfnamefont{G.~G.} \bibnamefont{Raffelt}},
  \bibinfo{journal}{Phys. Rev.} \textbf{\bibinfo{volume}{D70}},
  \bibinfo{pages}{043526} (\bibinfo{year}{2004}), \eprint{astro-ph/0403417}.

\bibitem[{\citenamefont{Ho and Scherrer}(2013{\natexlab{a}})}]{Ho:2012ug}
\bibinfo{author}{\bibfnamefont{C.~M.} \bibnamefont{Ho}} \bibnamefont{and}
  \bibinfo{author}{\bibfnamefont{R.~J.} \bibnamefont{Scherrer}},
  \bibinfo{journal}{Phys. Rev.} \textbf{\bibinfo{volume}{D87}},
  \bibinfo{pages}{023505} (\bibinfo{year}{2013}{\natexlab{a}}),
  \eprint{1208.4347}.

\bibitem[{\citenamefont{Ho and Scherrer}(2013{\natexlab{b}})}]{Ho:2012br}
\bibinfo{author}{\bibfnamefont{C.~M.} \bibnamefont{Ho}} \bibnamefont{and}
  \bibinfo{author}{\bibfnamefont{R.~J.} \bibnamefont{Scherrer}},
  \bibinfo{journal}{Phys. Rev.} \textbf{\bibinfo{volume}{D87}},
  \bibinfo{pages}{065016} (\bibinfo{year}{2013}{\natexlab{b}}),
  \eprint{1212.1689}.

\bibitem[{\citenamefont{Boehm et~al.}(2013)\citenamefont{Boehm, Dolan, and
  McCabe}}]{Boehm:2013jpa}
\bibinfo{author}{\bibfnamefont{C.}~\bibnamefont{Boehm}},
  \bibinfo{author}{\bibfnamefont{M.~J.} \bibnamefont{Dolan}}, \bibnamefont{and}
  \bibinfo{author}{\bibfnamefont{C.}~\bibnamefont{McCabe}},
  \bibinfo{journal}{JCAP} \textbf{\bibinfo{volume}{1308}}, \bibinfo{pages}{041}
  (\bibinfo{year}{2013}), \eprint{1303.6270}.

\bibitem[{\citenamefont{Steigman}(2013)}]{Steigman:2013yua}
\bibinfo{author}{\bibfnamefont{G.}~\bibnamefont{Steigman}},
  \bibinfo{journal}{Phys. Rev.} \textbf{\bibinfo{volume}{D87}},
  \bibinfo{pages}{103517} (\bibinfo{year}{2013}), \eprint{1303.0049}.

\bibitem[{\citenamefont{Nollett and Steigman}(2014)}]{Nollett:2013pwa}
\bibinfo{author}{\bibfnamefont{K.~M.} \bibnamefont{Nollett}} \bibnamefont{and}
  \bibinfo{author}{\bibfnamefont{G.}~\bibnamefont{Steigman}},
  \bibinfo{journal}{Phys. Rev.} \textbf{\bibinfo{volume}{D89}},
  \bibinfo{pages}{083508} (\bibinfo{year}{2014}), \eprint{1312.5725}.

\bibitem[{\citenamefont{Nollett and Steigman}(2015)}]{Nollett:2014lwa}
\bibinfo{author}{\bibfnamefont{K.~M.} \bibnamefont{Nollett}} \bibnamefont{and}
  \bibinfo{author}{\bibfnamefont{G.}~\bibnamefont{Steigman}},
  \bibinfo{journal}{Phys. Rev.} \textbf{\bibinfo{volume}{D91}},
  \bibinfo{pages}{083505} (\bibinfo{year}{2015}), \eprint{1411.6005}.

\bibitem[{\citenamefont{Dolgov}(2002)}]{Dolgov:2002wy}
\bibinfo{author}{\bibfnamefont{A.~D.} \bibnamefont{Dolgov}},
  \bibinfo{journal}{Phys. Rept.} \textbf{\bibinfo{volume}{370}},
  \bibinfo{pages}{333} (\bibinfo{year}{2002}), \eprint{hep-ph/0202122}.

\bibitem[{\citenamefont{Mangano et~al.}(2005)\citenamefont{Mangano, Miele,
  Pastor, Pinto, Pisanti, and Serpico}}]{Mangano:2005cc}
\bibinfo{author}{\bibfnamefont{G.}~\bibnamefont{Mangano}},
  \bibinfo{author}{\bibfnamefont{G.}~\bibnamefont{Miele}},
  \bibinfo{author}{\bibfnamefont{S.}~\bibnamefont{Pastor}},
  \bibinfo{author}{\bibfnamefont{T.}~\bibnamefont{Pinto}},
  \bibinfo{author}{\bibfnamefont{O.}~\bibnamefont{Pisanti}}, \bibnamefont{and}
  \bibinfo{author}{\bibfnamefont{P.~D.} \bibnamefont{Serpico}},
  \bibinfo{journal}{Nucl. Phys.} \textbf{\bibinfo{volume}{B729}},
  \bibinfo{pages}{221} (\bibinfo{year}{2005}), \eprint{hep-ph/0506164}.

\bibitem[{\citenamefont{de~Salas and Pastor}(2016)}]{deSalas:2016ztq}
\bibinfo{author}{\bibfnamefont{P.~F.} \bibnamefont{de~Salas}} \bibnamefont{and}
  \bibinfo{author}{\bibfnamefont{S.}~\bibnamefont{Pastor}},
  \bibinfo{journal}{JCAP} \textbf{\bibinfo{volume}{1607}}, \bibinfo{pages}{051}
  (\bibinfo{year}{2016}), \eprint{1606.06986}.

\bibitem[{\citenamefont{Cyburt et~al.}(2016)\citenamefont{Cyburt, Fields,
  Olive, and Yeh}}]{Cyburt:2015mya}
\bibinfo{author}{\bibfnamefont{R.~H.} \bibnamefont{Cyburt}},
  \bibinfo{author}{\bibfnamefont{B.~D.} \bibnamefont{Fields}},
  \bibinfo{author}{\bibfnamefont{K.~A.} \bibnamefont{Olive}}, \bibnamefont{and}
  \bibinfo{author}{\bibfnamefont{T.-H.} \bibnamefont{Yeh}},
  \bibinfo{journal}{Rev. Mod. Phys.} \textbf{\bibinfo{volume}{88}},
  \bibinfo{pages}{015004} (\bibinfo{year}{2016}), \eprint{1505.01076}.

\bibitem[{\citenamefont{Arbey}(2012)}]{Arbey:2011nf}
\bibinfo{author}{\bibfnamefont{A.}~\bibnamefont{Arbey}},
  \bibinfo{journal}{Comput. Phys. Commun.} \textbf{\bibinfo{volume}{183}},
  \bibinfo{pages}{1822} (\bibinfo{year}{2012}), \eprint{1106.1363}.

\bibitem[{\citenamefont{Hickerson}(2016 (accessed March 2,
  2016))}]{Hickerson:2016BBNCODE}
\bibinfo{author}{\bibfnamefont{K.}~\bibnamefont{Hickerson}},
  \emph{\bibinfo{title}{{AlterBBN version 1.6}}} (\bibinfo{year}{2016 (accessed
  March 2, 2016)}), \urlprefix\url{https://github.com/hickerson/bbn}.

\bibitem[{\citenamefont{Galvez}(2016)}]{Galvez:2016BBNCODE}
\bibinfo{author}{\bibfnamefont{R.}~\bibnamefont{Galvez}},
  \emph{\bibinfo{title}{{BBN variant to allow varying neutrino number and
  neutrino temperature}}} (\bibinfo{year}{2016}),
  \urlprefix\url{https://github.com/richardagalvez/BBN_varying_neutrino}.

\bibitem[{\citenamefont{Olive et~al.}(2014)}]{Agashe:2014kda}
\bibinfo{author}{\bibfnamefont{K.~A.} \bibnamefont{Olive}} \bibnamefont{et~al.}
  (\bibinfo{collaboration}{Particle Data Group}), \bibinfo{journal}{Chin.
  Phys.} \textbf{\bibinfo{volume}{C38}}, \bibinfo{pages}{090001}
  (\bibinfo{year}{2014}).

\bibitem[{\citenamefont{Bratt et~al.}(2002)\citenamefont{Bratt, Gault,
  Scherrer, and Walker}}]{Bratt:2002xt}
\bibinfo{author}{\bibfnamefont{J.~D.} \bibnamefont{Bratt}},
  \bibinfo{author}{\bibfnamefont{A.~C.} \bibnamefont{Gault}},
  \bibinfo{author}{\bibfnamefont{R.~J.} \bibnamefont{Scherrer}},
  \bibnamefont{and} \bibinfo{author}{\bibfnamefont{T.~P.}
  \bibnamefont{Walker}}, \bibinfo{journal}{Phys. Lett.}
  \textbf{\bibinfo{volume}{B546}}, \bibinfo{pages}{19} (\bibinfo{year}{2002}),
  \eprint{astro-ph/0208133}.

\bibitem[{\citenamefont{Cooke et~al.}(2016)\citenamefont{Cooke, Pettini,
  Nollett, and Jorgenson}}]{Cooke:2016rky}
\bibinfo{author}{\bibfnamefont{R.}~\bibnamefont{Cooke}},
  \bibinfo{author}{\bibfnamefont{M.}~\bibnamefont{Pettini}},
  \bibinfo{author}{\bibfnamefont{K.~M.} \bibnamefont{Nollett}},
  \bibnamefont{and}
  \bibinfo{author}{\bibfnamefont{R.}~\bibnamefont{Jorgenson}},
  \bibinfo{journal}{Astrophys. J.} \textbf{\bibinfo{volume}{830}},
  \bibinfo{pages}{148} (\bibinfo{year}{2016}), \eprint{1607.03900}.

\bibitem[{\citenamefont{Izotov et~al.}(2014)\citenamefont{Izotov, Thuan, and
  Guseva}}]{Izotov:2014fga}
\bibinfo{author}{\bibfnamefont{Y.~I.} \bibnamefont{Izotov}},
  \bibinfo{author}{\bibfnamefont{T.~X.} \bibnamefont{Thuan}}, \bibnamefont{and}
  \bibinfo{author}{\bibfnamefont{N.~G.} \bibnamefont{Guseva}},
  \bibinfo{journal}{Mon. Not. Roy. Astron. Soc.}
  \textbf{\bibinfo{volume}{445}}, \bibinfo{pages}{778} (\bibinfo{year}{2014}),
  \eprint{1408.6953}.

\bibitem[{\citenamefont{Aver et~al.}(2015)\citenamefont{Aver, Olive, and
  Skillman}}]{Aver:2015iza}
\bibinfo{author}{\bibfnamefont{E.}~\bibnamefont{Aver}},
  \bibinfo{author}{\bibfnamefont{K.~A.} \bibnamefont{Olive}}, \bibnamefont{and}
  \bibinfo{author}{\bibfnamefont{E.~D.} \bibnamefont{Skillman}},
  \bibinfo{journal}{JCAP} \textbf{\bibinfo{volume}{1507}}, \bibinfo{pages}{011}
  (\bibinfo{year}{2015}), \eprint{1503.08146}.

\bibitem[{\citenamefont{Fields}(2011)}]{Fields:2011zzb}
\bibinfo{author}{\bibfnamefont{B.~D.} \bibnamefont{Fields}},
  \bibinfo{journal}{Ann. Rev. Nucl. Part. Sci.} \textbf{\bibinfo{volume}{61}},
  \bibinfo{pages}{47} (\bibinfo{year}{2011}), \eprint{1203.3551}.

\end{thebibliography}
\end{document}